\documentstyle[aaspp4]{article}
\begin{document}
\title{\sc Optical Spectrum of  Main--, Inter-- and Off-pulse 
Emission from Crab Pulsar}

\author{Alberto Carrami\~nana\altaffilmark{1}, 
Andrej \v Cade\v z\altaffilmark{2},
Toma\v z Zwitter\altaffilmark{2}}

\altaffiltext{1}{I.N.A.O.E., Luis Enrique Erro 1, Tonantzintla, Puebla 72840, M\'exico}
\altaffiltext{2}{Department of Physics, University of Ljubljana, Jadranska 19, 
1000 Ljubljana, Slovenia}

\received{\ } 
\accepted{ for publication in Astrophysical Journal on May 22, 2000}

\begin{abstract}
A dedicated stroboscopic device was used to obtain 
optical spectra of the Crab main--pulse and inter--pulse
as well as the spectrum of the underlying 
nebula when the pulsar is turned off. 
As the nebular emission is very inhomogeneous, our ability to effectively
subtract the nebular background signal is crucial.

No spectral lines intrinsic to the pulsar are detected. The main-pulse
and the inter-pulse behave as power laws, both with the same de-reddened index  
$\alpha = +0.2 \pm 0.1$.
This value was obtained by subtracting the nebular spectrum at the exact 
position of the pulsar. The underlying nebula is redder, 
$\alpha = -0.4 \pm 0.1$. Its emission lines are split into approaching 
($\sim -1200$~km/s) and receding ($\sim +600$~km/s) components. The strength 
of emission line components and the flux in nebular continuum 
vary on arcsec scale. The nebular line and continuum intensities 
along the N-S slit are given.
\end{abstract}

\keywords{Pulsars: individual (PSR B0531$+$21) ---
stars: neutron --- techniques: spectroscopic  --- supernova remnants}

\section{Introduction}

The Crab pulsar (PSR J0534+2200 = PSR B0531$+$21) was the first pulsating 
radio source identified at optical wavelengths (Cocke, Disney, \&\ Taylor 
1969; Lynds, Maran, \&\ Trumbo 1969). 
Most subsequent studies concentrated on the pulse shape in various 
photometric bands (Percival et al. 1993; Eikenberry et al.\ 1996; 
Perryman et al.\ 1999; Lundqvist et al.\ 1999). 
They generally suggested small color variations in the 
light curve. No spectroscopic data existed until Nasuti et al.\ (1996) 
obtained the first optical spectrum. They were able to measure the general 
flux distribution as well as hint to an unidentified intrinsic absorption 
feature at 5920~\AA. Their spectrum is phase averaged, so no information on 
spectral variation with pulse phase can be extracted. 

Here we present the first flux-calibrated optical spectrum, separately for the 
main-pulse, the inter-pulse and the off-pulse components. The spectra of both
pulses are compared with phase-resolved photometric studies. 
Except for a very small ($\sim 1$\% ) contribution from the pulsar 
(Golden et al.\ 2000), the off-pulse spectrum, also presented here, 
corresponds to the nebular emission along the line of sight of the pulsar.
 
\section{Observations}

Observations were obtained with the LFOSC spectrograph at the 2.12m telescope 
of the Observatorio Astrof\'{\i}sico Guillermo Haro in Cananea, Mexico. 
The useful wavelength range was $\lambda\lambda 4370 - 9000 $~\AA\ and the 
spectral resolution was $\sim 15$~\AA. The slit of 2 arcsec width was kept 
in the N--S direction. 

The light entered the spectroscope through a frequency and phase controlled 
chopper. The device, described by \v{C}ade\v{z} \&\  Gali\v{c}i\v{c} (1996), 
was used before on the same telescope to study 
the optical light curve of the pulsar. A blade allows light to enter 
the spectrograph for 10\%\ of each pulse period. This collection window 
is synchronized with the pulsar and kept on the desired pulse phase interval. 
We use the most recent of the Jodrell-Bank monthly Crab ephemerides, 
translated from the Solar-System Barycentre, including real-time corrections 
for the Earth orbital motion. 
Long photometric observations confirmed that the difference between the 
pulsar and the chopper phase remains within $5 \times 10^{-4}$ for at least 
3 hours (\v Cade\v z et al.\ 2000). However the initial 
chopper--pulsar phase is set by maximizing the apparent magnitude of 
the pulsar, so that the absolute phase is only accurate to 0.03 
(Gali\v ci\v c 1999).

Observations were obtained on Dec 13 and Dec 14 1998 and on Oct 18 and Oct 20 
1999 under nearly photometric conditions and $\sim 1"$ seeing. 
A total accumulated exposure on the main-pulse and the inter-pulse was 
11700~sec and 9000~sec, while the total off-pulse observing time 
was 10200~sec. 
The off-pulse interval was chosen $\simeq 0.25$ in phase after the main-pulse. 
Several direct images of the pulsar field were obtained in order to control 
the atmosphere
transparency and phase of the pulsar. The wavelength calibration is generally 
good to 2~\AA\ and somewhat deteriorating at the blue end. The 
flux calibration, obtained through several exposures of the star EG50 and 
Hiltner~600, is accurate to 5\%\ in both color and absolute fluxes.  
All spectra were obtained between airmasses 1.01 and 1.3. 
The pulsar spectra were de-reddened assuming $E_{B-V}= A_V/3.1 = 0.51$ 
(Percival et al.\ 1993) and
\begin{equation}
\label{reden}
A_\lambda / E_{B-V} = -7.51 \log(\lambda/\lambda_o)+1.15;\ \ \   
\lambda_o = 1\mu{\rm m} 
\end{equation}
which follows the average 
Galactic extinction curve (Savage \&\ Mathis 1979) to better than 1\%\ 
for $0.48<\lambda/\lambda_o<1.0$. 

\begin{table}
\caption[]{ \label{alphas} Parametrization of fluxes in the 0.1-phase 
window centered on the main- and inter-pulse. The last line is 
unpulsed flux of the underlying nebular continuum in the 1 arsec$^2$ box 
centered on the pulsar. 
$F_\lambda = K \times (\lambda/\lambda_o) ^ {-\alpha-2}$,    
$\lambda_o = 6000$~\AA, 5000\AA~$< \lambda < 7500$\AA. 
Fluxes are in erg s$^{-1}$/cm$^2$/\AA. All data were de-reddened using
eq.\ \ref{reden}.
\medskip \ }
\begin{center}
\begin{tabular}{lrr}
\tableline \tableline 
                   & $K$                  & $\alpha$ \\ \tableline
main-pulse       & $5.9 \times 10^{-15}$ & $ +0.2 \pm 0.1$ \\
inter-pulse      & $1.9 \times 1
0^{-15}$ & $ +0.2 \pm 0.1$ \\
underlying nebula& $3.4 \times 10^{-15}$ & $ -0.4 \pm 0.1$ \\
\tableline
\end{tabular}
\end{center}
\end{table}

Spectra were reduced with standard IRAF routines used according to our 
particular purpose. First, all the spectra were aligned 
with respect to the tracing 
of the pulsar and of another star 11.4 arcsec South of the pulsar. 
In order to achieve an optimal flux calibration, a relatively wide 
aperture of 7 arcsec was used that included all pulsar-related flux and 
allowed for any wavelength variations of the focus of the spectrograph. 
All individual spectra were flux-calibrated but no nebular background was 
subtracted at this stage. 
The un-pulsed light was subtracted subsequently by subtracting the 
off-pulse flux-calibrated spectrum obtained just before or after.    
This technique makes an assumption-free subtraction of any nebular 
light, scattered light or telluric contributions. 
The chopper effectively enables us to take the pulsar off the sky and subtract
the nebula emission at the exact pulsar's position. This is important as the
nebular emission is varying on arcsec scale (cf.\ Section \ref{secnebula}).

\section{Pulsar}

The flux-calibrated de-reddened spectra of the main-pulse and inter-pulse 
are presented in the upper panel of Fig.\ 1. The surface flux of the 
underlying nebula is plotted in the lower panel. Both axes are plotted in 
logarithmic scale. 

\begin{table}
\caption[]{ Upper limits to de-reddened \label{lines} 
fluxes (in erg s$^{-1}$ cm$^{-2}$) 
of nebular emission lines for the 0.1-phase window centered on the main- and 
the  inter- pulse. The last column gives fluxes of nebular lines in a 
2 arcsec (E-W) $\times 6.9 $~arcsec (N-S) box centered on the pulsar. 
\medskip \ }
\begin{center}
\begin{tabular}{lrrr}
\tableline \tableline 
     & main-pulse & inter-pulse & nebula \\ \tableline
$[$ O III $]$  4959+5007 & 
$<1.0\times 10^{-13}$  & $<8.2\times 10^{-14}$   & $8.4\times 10^{-12}$\\
$[$ O I $]$   6300 & 
$<1.3\times 10^{-14}$  & $<1.2\times 10^{-14}$   & $8.6\times 10^{-13}$\\
$[$ O I $]$   6363 & 
$<9.5\times 10^{-15}$  & $<8.8\times 10^{-15}$   & $2.6\times 10^{-13}$\\
H I + $[$ N II $]$ 6548-84 & 
$<2.5\times 10^{-14}$  & $<3.0\times 10^{-14}$   & $4.5\times 10^{-12}$\\
$[$ S II $]$  6716+6731& 
$<2.2\times 10^{-14}$  & $<2.7\times 10^{-14}$   & $2.4\times 10^{-12}$\\
\tableline
\end{tabular}
\end{center}
\end{table}

The colors of the main-pulse and inter-pulse continuum are identical,
but the nebular continuum is redder (see Table \ref{alphas}). 

The pulsar's spectrum features no lines apart from telluric bands and 
weak nebular line residuals. Note that the $[$OI$]$ 5579 line is of 
telluric origin. The upper limits to equivalent widths of nebular 
lines are 5 and 10 \AA\ for the main- and inter-pulse respectively. 
This translates to the very stringent limits on absolute fluxes of pulsed 
emission lines that are given in Table \ref{lines}.   

We did not find any trace of the 5920~\AA\ absorption feature hinted 
by Nasuti et al.\ (1996). 
We conclude that they were right in attributing this feature to imperfect 
flux calibration of their spectrum. 

Note that the nebular continuum is rather strong. The use of 
the 0.1-phase window with the chopper however suppressed it by a 
factor of 10. This emphasizes the advantage of our observing method. 

\section{Nebula\label{secnebula}}

Davidson (1979) published the first study of selected condensations in the Crab
nebula. Nasuti et al.\  (1996) discuss the nebular spectrum 2 arcsec East
of the pulsar position. Our slit had a N--S orientation so a brief discussion of
the spectrum along this direction as well as directly on the pulsar position is
in order.

\begin{table}
\caption[]{ \label{tab} Radial velocities (in km/s) of nebular lines 
within 8 arc sec of the pulsar.\medskip \ }
\begin{center}
\begin{tabular}{lclcrcl}
\tableline \tableline 
&
\multicolumn{3}{c}{Approaching}&\multicolumn{3}{c}{Receding} \\ \tableline
$[$ O III $]$ 4959, 5007 & --1200 &$\pm$& 100 & $+$1150 &$\pm$& 100 \\
$[$ O I $]$   6300, 6363 & --1050 &$\pm$& 200 & $+$600  &$\pm$& 200 \\
H I 6563 &       --1200 &$\pm$& 100 & $+$650  &$\pm$& 100 \\
$[$ N II $]$  6548, 6584 & --1200 &$\pm$& 100 & $+$650  &$\pm$& 150 \\
$[$ S II $]$  6716, 6731 & --1300 &$\pm$& 200 & $+$650  &$\pm$& 200  \\ 
\tableline
\end{tabular}
\end{center}

\end{table}

The shape of the (de-reddened) nebular continuum is constant along the slit
($\alpha = -0.4\pm 0.1$) except for its strength. Spatial variations of 
surface flux continuum (de-reddened nebular continuum per square arc-sec) are 
shown in the lower panel of Fig.~2. The horizontal axis is the relative 
displacement in N-S direction with respect to the position of 
the pulsar and the vertical axis displays the continuum surface flux 
at 6000\AA . There is a chance superposition 
of a star 11.4 arcsec South of the pulsar, so no 
flux is given around its position.

The upper panels of Fig.~2 give de-reddened surface fluxes of emission lines as 
a function of position relative to the pulsar. The lines are generally 
well-separated into approaching and receding components (Fig.~3). 
The exception is a 
blend of the red component of [O~III] 4959 and the blue component of its 5007 
counterpart which are plotted together. The H~I 6563, [N~II] 6548+6584 region 
was resolved assuming a standard 3 to 1 ratio of the [N~II] line intensities. 

All lines show well separated red and blue components, similar to those 
reported by Nasuti et al.\  (1996). Component velocities, given in Table 
\ref{tab}, remain constant (to within 
200 km/s) up to $\sim 30$~arcsec from the pulsar. All lines show similar 
velocity shifts, $\sim 1800 \rm km/s$, except for the very strong receding 
component of [O~III] 4959 and 5007 lines, receding at double velocity. 

The intensity of blue and red components is highly spatially variable and 
uncorrelated (Fig.~2). A simple interpretation is that of an expanding shell,
proposed already by Davidson (1979).
The receding components of forbidden nebular lines show a wide pronounced peak
$\sim 3$~arcsec North of the pulsar. The slope of the condensation, however, 
extends well behind the pulsar itself. The approaching components are generally
weaker and show a condensation $\sim 4$~arcsec South of the pulsar. 
The kinematics and strength of  H$\alpha$ is similar to that of forbidden 
lines, but a pronounced receding component to the North of the pulsar is 
missing.

\section{Discussion and conclusions}

The main result of this work is the flux calibrated time resolved spectrum 
of the Crab pulsar. It shows no intrinsic spectral lines with the continuum
following a power law with $\alpha = +0.2 \pm 0.1$. The spectra of the 
main-pulse and the inter-pulse are identical: flux of the main-pulse 
multiplied by 0.32 equals that of the inter-pulse  
(Fig.\ 4), where the formal upper limit to the difference in the power law 
index for the main- and the inter-pulse is 0.01 ($2\sigma$).

The above value of $\alpha$ refers to de-reddened main- and inter-pulse
spectra with nebular spectra subtracted at the exact area of pulsar's 
PSF. This crucial subtraction was made possible by the dedicated 
chopper device that effectively enabled us to take the pulsar off the 
sky. 

The underlying nebula is redder ($\alpha = -0.4 \pm 0.1$) and of varying 
intensity. 
We expect that this fact may explain why the power law index for the 
pulsar spectrum reported in the literature is slightly redder than deduced 
here. Nasuti et al.\  (1996) find
$\alpha = -0.1\pm 0.01$ in optical. Gull et al. (1998) give 
$\alpha = -0.1\pm 0.2$ (if $E_{B-V} = 0.51$ is assumed) in the UV.
Without the possibility of directly subtracting the off-pulse spectrum, 
one must make a model for the surrounding nebula. We simulated such 
modeling and took in an example the nebular spectrum 4 arcsec North and 
South from the pulsar as the background. This led to the derived value 
of $\alpha = +0.1$ in very good seeing (FWHM $\sim 0.7$ arcsec) or even 
0.0 in more mediocre conditions. We also note that errors below 0.1 in 
$\alpha$ 
appear unrealistic, as they imply a color calibration accurate to better than 
7\% -- a goal difficult to achieve on a patchy nebular background and 
given the uncertainties in the extinction law.   

We detected no color variation in the pulsed light. Photometric 
investigation of Eikenberry et al.\ (1996) reports slight color 
variation on pulse slopes. Note, however, that our observations were 
obtained close to the 
centers of the main- and inter-pulse. So the two studies 
are somewhat difficult to compare: we present flux-calibrated spectra with 
slightly uncertain absolute phase, while Eikenberry et al.\ have an 
excellent phase definition but lack absolute color calibration. 
Romani et al. (1999) report on a relative decrease of the 
main pulse flux in the IR ($1.4 \mu$m~$< \lambda < 1.8 \mu$m) with respect to 
the optical ($0.38 \mu$m~$< \lambda < 0.83 \mu$m). 
We can not confirm a similar trend in our optical spectra, even if the 
braking of the power law beyond
800~nm could possibly be interpreted in this way. Since the nebular 
spectrum has a similar break, we would like another independent confirmation. 
A phase resolved spectrum of the Crab was presented by Perryman et al.~(1999).
In agreement with our results they detected no color dependence of the 
pulse shape in the optical. 
Note however that a limited resolution ($\sim 100\rm nm$) and performance 
of the otherwise promising STJ junction detectors -- together with modest 
seeing -- prevented them from flux calibrating their spectrum and subtracting 
the nebular contributions. 

By perfecting the stroboscopic technique we were able to obtain 
the spectrum of the pulsed emission of the Crab pulsar on a medium size
telescope. We are now using the same technique to identify optical 
counterparts of known radio pulsars.

\acknowledgements

A.\v{C}. and T.Z. acknowledge support from the Slovenian Ministry for 
Science and Technology. We would like to thank an anonymous referee 
whose comments improved the presentation of this paper.


\begin{figure}
\figurenum{1}
\plotone{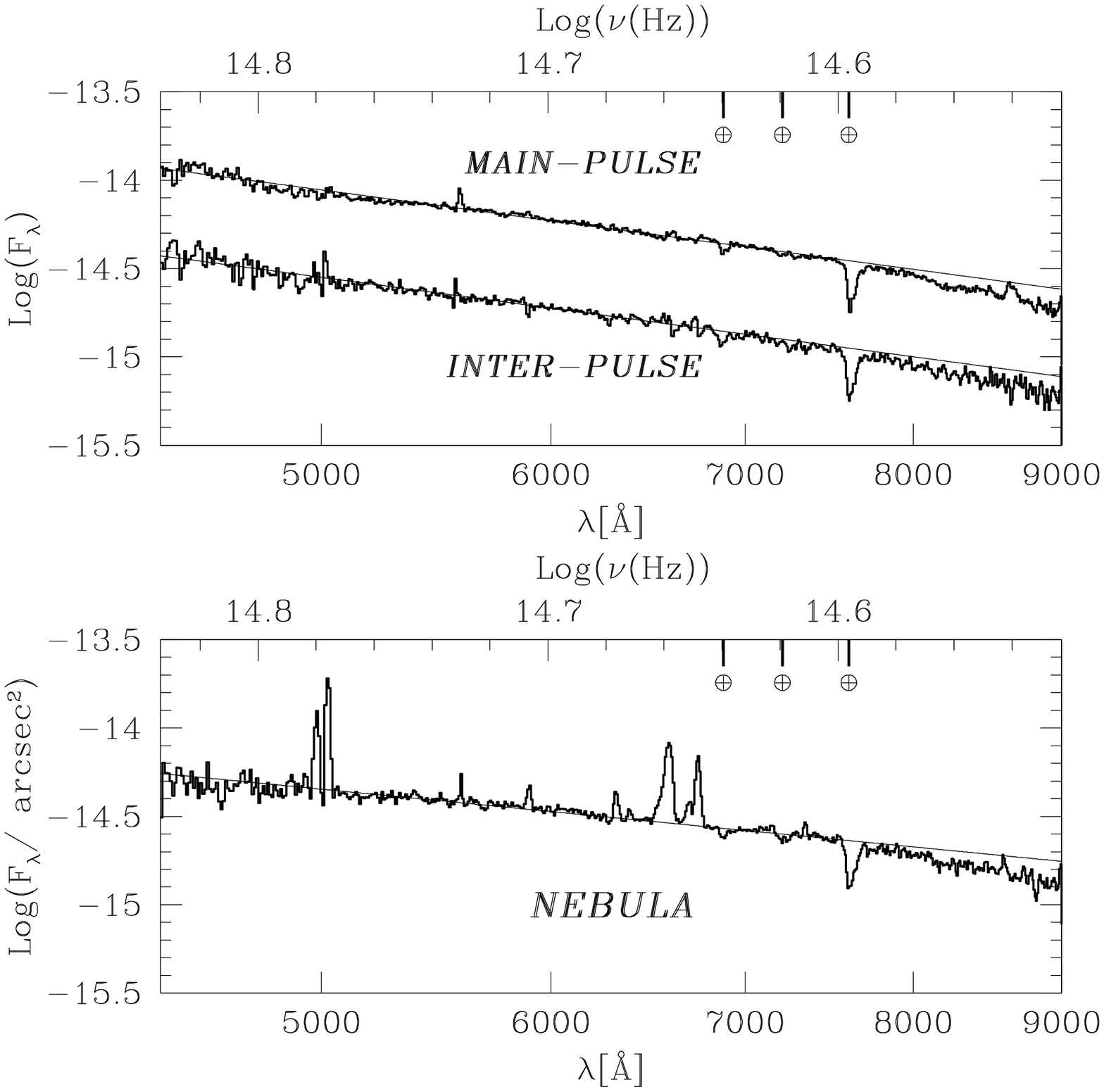}
\caption[1]{{\em Upper panel:} Log($F_\lambda$) - Log($\lambda$) 
spectra of the main--pulse and the inter--pulse. Emission within a 
0.1 phase-window, centered on pulse's peak, is shown. Flux is in 
erg~s$^{-1}$~cm$^{-2}$~\AA$^{-1}$. Straight lines are power law fits
(cf. Table \ref{reden}). \\
{\em Lower panel:} 
The flux in the nebula within 1~arcsec$^2$ box centered on the pulsar. 
}

\end{figure}

\begin{figure}
\figurenum{2} \plotone{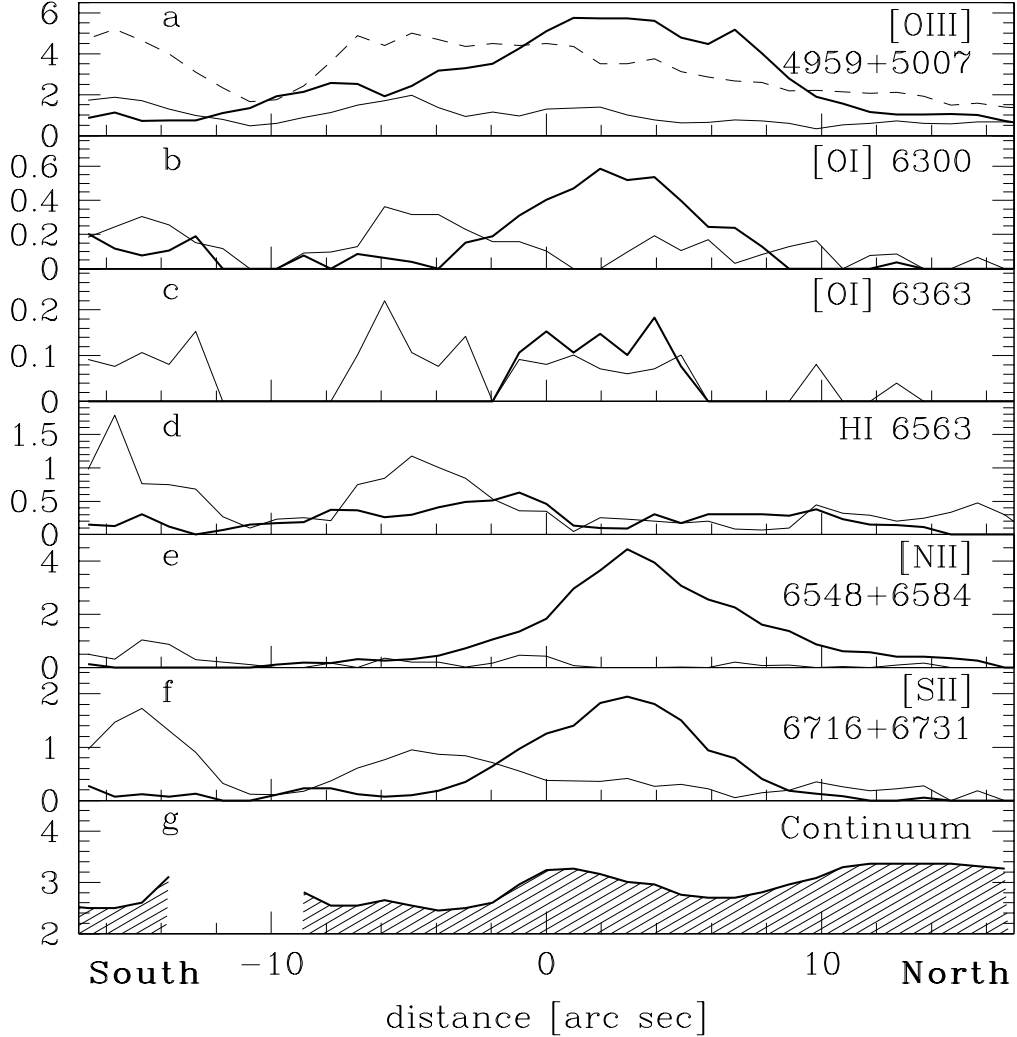}
\caption[2]{{\em Panels (a--f):} 
Spatial variation of fluxes of the nebular emission lines. Flux is in 
$10^{-13}$~erg~s$^{-1}$~cm$^{-2}$~arcsec$^{-2}$. 
The slit had 
a N-S orientation, distance along the slit is given with respect to the 
position of the pulsar. {\it Bold lines} refer to a receding and {\it normal 
lines} to an approaching component of each line. Components of [OIII] lines 
are overlapping; 
so we plot the receding component of 5007 [bold], the approaching of 4959
[normal] and the sum of 5007 (blueshifted) and 4959 (redshifted) [dashed line].
Flux errors are $\sim 0.15 \times 10^{-13}$~erg~s$^{-1}$~cm$^{-2}$~arcsec$^{-2}$
($2 \sigma$), rendering them small except for the [OI] lines.
\\ 
{\em Panel (g):} Flux in the continuum at 6000\AA\ in units of 
$10^{-15}$~erg~s$^{-1}$~cm$^{-2}$~\AA$^{-1}$~arcsec$^{-2}$. Region 
$\sim 11$~arcsec south of the pulsar is omitted due to contribution 
of a chance superposition star.
 } 
 
\end{figure}

\begin{figure}
\figurenum{3}
\plotfiddle{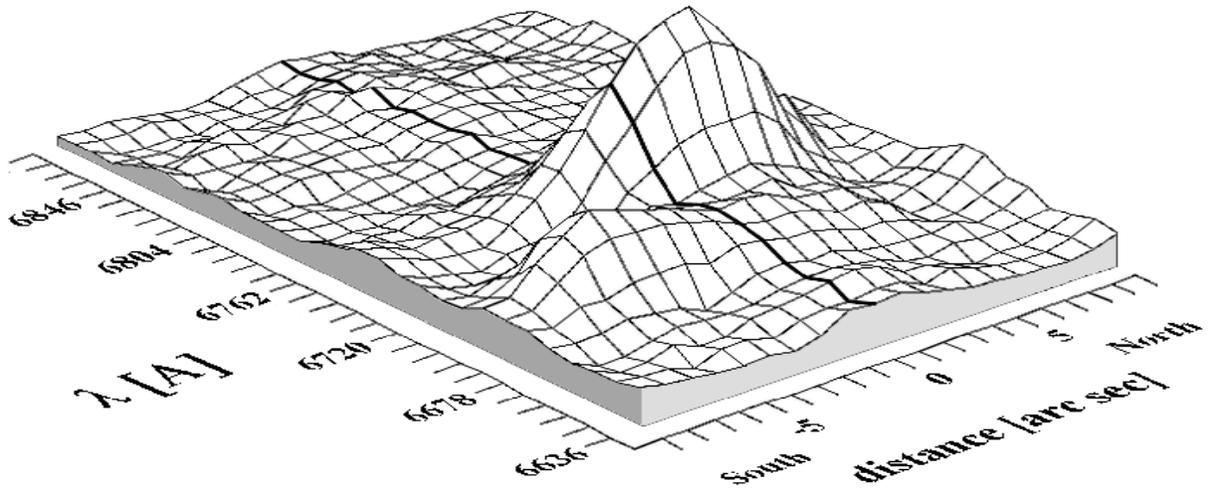}{3.2in}{-90}{65.0}{50.0}{-270}{300} 
\caption[3]{
Portion of the flux calibrated spectrum of the nebula obtained with the 
pulsar turned off (phase $\simeq 0.25$ after the main--pulse peak). The 
strong blue- and red-shifted components of the S[II]~6716$+$6731 doublet 
are seen. Their intensity varies on arc-sec scale, but their radial 
velocity remains constant. Thick line marks the position of the pulsar.
}
\end{figure}

\newpage

\begin{figure}
\figurenum{4}
\plotfiddle{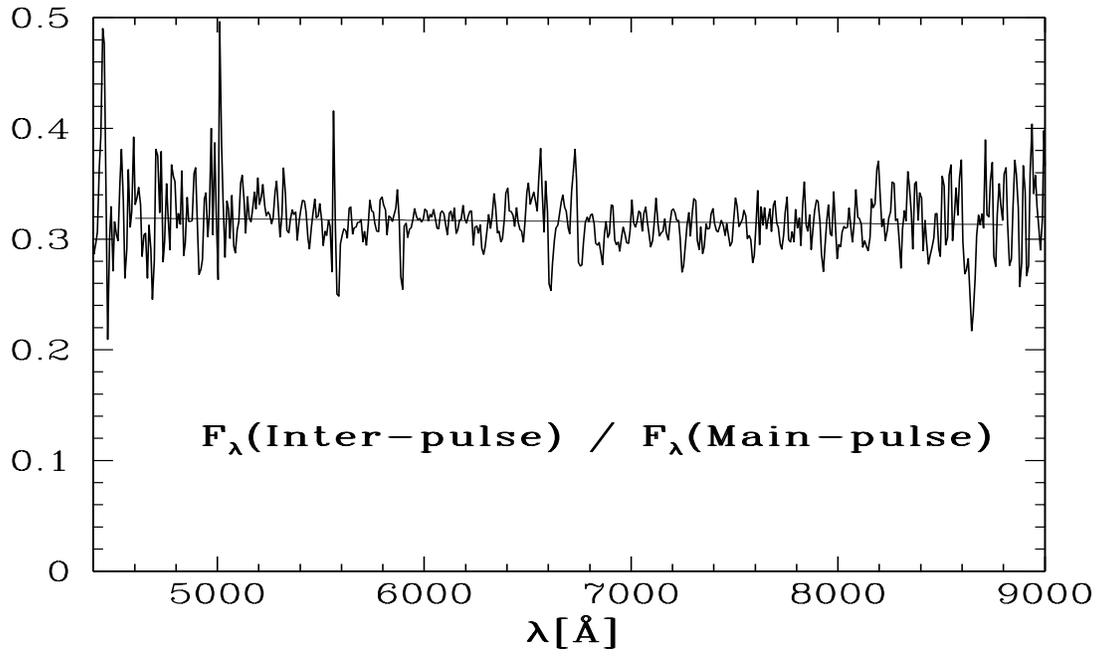}{3.0in}{-90}{60.0}{45.0}{-250}{300}
\caption[4]{
Ratio of the inter--pulse and main--pulse spectra. It is well represented
with a linear fit 
$0.317 (1 \pm 0.003) - 0.006 (1 \pm 1) (\lambda / \lambda_o -1)$;
$\lambda_o = 6000$~\AA .
Thus the spectral shapes of the main and inter-pulse in the 
4400 \AA~$< \lambda < 8800$~\AA\  wavelength interval are identical. 
}
\end{figure}

\end{document}